\documentclass[10pt]{article}
\usepackage{latexsym,graphicx}
\newcommand{\be}{\begin{equation}}
\newcommand{\ee}{\end{equation}}
\def\n{\noindent}
\catcode `\@=11
\catcode `\@=12

\begin{document}
\begin{center}
\large{\bf { GENERATION OF BIANCHI TYPE V BULK VISCOUS COSMOLOGICAL MODELS WITH TIME DEPENDENT 
$\Lambda$-TERM }}\\ 
\vspace{10mm}
\normalsize{ANIRUDH PRADHAN$^a$, KANTI JOTANIA$^b$ and ANJU RAI$^c$} \\
\vspace{5mm}
\normalsize{$^a$ {\it Dept. of Mathematics, Hindu Post-graduate College,
 Zamania-232 331, Ghazipur, India} \\
\normalsize{E-mail address: pradhan@iucaa.ernet.in}}\\
\vspace{5mm}
\normalsize{$^b$ {\it Dept. of Physics, Faculty of Science, The M.S. University of Baroda, 
Vadodara-390 002, Gujarat, India} \\
\normalsize{E-mail address: kanti@iucaa.ernet.in}}\\
\vspace{5mm}
\normalsize{$^c$ {\it Dept. of Mathematics, Hindu Post-graduate College, Zamania 232 331,
Ghazipur, India} \\ 
\normalsize{E-mail address: anjuraianju@yahoo.co.in}}
\end{center}
\vspace{10mm}
%\date{}
%\maketitle
\begin{abstract} 
Bianchi type V bulk viscous fluid cosmological models are investigated with 
dynamic cosmological term $\Lambda(t)$. Using a generation 
technique (Camci {\it et al.}, 2001), it is shown that the Einstein's field 
equations are solvable for any arbitrary cosmic scale function. Solutions for 
particular forms of cosmic scale functions are also obtained. The cosmological 
constant is found to be decreasing function of time, which is supported by 
results from recent type Ia supernovae observations. Some physical and geometrical 
aspects of the models are also discussed.  
\end{abstract}
\smallskip
%\n PACS number: {98.80.Es, 98.80.-k}  \\
\n Keywords: {cosmology, Bianch type V universe, cosmological constant}

%%%%%%%%%%%%%%%%%%%%%%%%%%%%%%%%%%%%%%%%%%%%%%%%%%%%%%%%%%%%%%%%%%%%%%%%%%%%%%%%%%%
%%%%%%%%%%%%%%%%%%%%%%%%%%%%%%%   SECTION 1  %%%%%%%%%%%%%%%%%%%%%%%%%%%%%%%%%%%%%%
\section{Introduction}
The study of Bianchi type V cosmological models create more interest as these 
models contain isotropic special cases and permit arbitrary small anisotropy 
levels at some instant of cosmic time. This property makes them suitable as
model of our universe. The homogeneous and isotropic Friedman-Robertson-Walker
(FRW) cosmological models, which are used to describe standard cosmological models,
are particular case of Bianchi type I, V and IX universes, according to whether 
the constant curvature of the physical three-space, $t$ = constant, is zero, 
negative or positive. These models will be interesting to construct cosmological
models of the types which are of class one. Present cosmology is based on the
FRW model which is completely homogeneous and isotropic. This is in agreement with 
observational data about the large scale structure of the universe. However, although 
homogeneous but anisotropic models are more restricted than the inhomogeneous models,
they explain a number of observed phenomena quite satisfactorily. This stimulates the 
research for obtaining exact anisotropic solution for Einstein's field equations 
(EFEs) as a cosmologically accepted physical models for the universe (at least in 
the early stages). Roy and Prasad \cite{ref1} have investigated Bianchi type V 
universes which are locally rotationally symmetric and are of embedding class one 
filled with perfect fluid with heat conduction and radiation. Bianchi type V 
cosmological models have been studied by other researchers (Farnsworth \cite{ref2} , 
Maartens and Nel \cite{ref3}, Wainwright {\it et al.} \cite{ref4}, Collins \cite{ref5}, 
Meena and Bali \cite{ref6}, Pradhan {\it et al.} \cite{ref7,ref8}) in different context. 

Models with a dynamic cosmological term $\Lambda(t)$ are becoming popular as they 
solve the cosmological-constant problem in a natural way. There is a significant 
observational evidence for the detection of Einstein's cosmological constant, $\Lambda$
or a component of material content of the universe, that varies slowly with time and 
space and so acts like $\Lambda$. Recent cosmological observations by High-z Supernova 
Team and Supernova Cosmological Project (Garnavich {\it et al.} \cite{ref9}, Perlmutter 
{\it et al.} \cite{ref10}, Riess {\it et al.} \cite{ref11}, Schmidt {\it et al.} \cite{ref12}) 
strongly favour a significant and positive $\Lambda$ with the magnitude 
$\Lambda(G\hbar / c^{3}) \approx 10^{-123}$. These observations on magnitudes and red-shift 
of type Ia supernova suggest that our universe may be accelerating one with a large function 
of the cosmological density in the form of cosmological $\Lambda$-term. Earlier researches 
on this topic, are contained in Lodovico \cite{ref13}, Weinberg \cite{ref14}, 
Dolgov \cite{ref15} $-$ \cite{ref17}, Bertolami \cite{ref18}, Ratra and Peebles \cite{ref19}, 
Carrol, Press and Turner \cite{ref20}. Some of the recent discussions on the cosmological-constant 
``problem'' and consequence on cosmology with a time-varying cosmological-constant have been 
discussed by Tsagas and Maartens \cite{ref21}, Sahni and Starobinsky \cite{ref22}, 
Peeble \cite{ref23}, Padmanabhan \cite{ref24}, Vishwakarma \cite{ref25}, and Pradhan 
{\it et al.} \cite{ref26}. This motivates us to study the cosmological models, where $\Lambda$ 
varies with time. 

The distribution of matter can be satisfactorily described by a perfect fluid
due to the large scale distribution of galaxies in our universe. However, 
observed physical phenomena such as the large entropy per baryon and the 
remarkable degree of isotropy of the cosmic microwave background radiation, 
suggest analysis of dissipative effects in cosmology. Furthermore, there are
several processes which are expected to give rise to viscous effects. These
are the decoupling of neutrinos during the radiation era and the decoupling
of radiation and matter during the recombination era. Bulk viscosity is
associated with the GUT phase transition and string creation. Misner \cite{ref27}
has studied the effect of viscosity on the evolution of cosmological models. The
role of viscosity in cosmology has been investigated by Weinberg \cite{ref28}. Nightingale
\cite{ref29}, Heller and Klimek \cite{ref30} have obtained a viscous universes without initial
singularity. The model studied by Murphy \cite{ref31} possessed an interesting feature in 
which  big bang type of singularity of infinite space-time curvature does not occur 
to be a finite past. However, the relationship assumed by Murphy between the
viscosity coefficient and the matter density is not acceptable at large 
density. Thus, we should consider the presence of material distribution other than 
a perfect fluid to obtain a realistic cosmological models (see Gr$\o$n \cite{ref32} for 
a review on cosmological models with bulk viscosity). The effect of bulk viscosity on 
the cosmological evolution has been investigated by a number of authors in the framework 
of general theory of relativity. This motivates to study cosmological bulk viscous fluid model.
 
In recent years, several authors (Hajj-Boutros \cite{ref33}, Hajj-Boutros and Sfeila \cite{ref34},
Ram \cite{ref35}, Mazumder \cite{ref36} and Pradhan and Kumar \cite{ref37}) have investigated 
the solutions of EFEs for homogeneous but anisotropic models by using some different generation 
techniques. Bianchi spaces $I-IX$ are useful tools in constructing models of spatially 
homogeneous cosmologies (Ellis and MacCallum \cite{ref38}, Ryan and Shepley \cite{ref39}). 
From these models, homogeneous Bianchi type V universes are the natural generalization of the 
open FRW model which eventually isotropize. Recently Camci {\it et al.} \cite{ref40}
derived a new technique for generating exact solutions of EFEs with perfect fluid 
for Bianchi type V space-time. Very recently Pradhan {\it et al.} \cite{ref41} have obtained 
Bianchi type V perfect fluid cosmological models with time dependent $\Lambda$-term.

In this paper, in what follows, we will discuss Bianchi type V cosmological models obtained 
by augmenting the energy-momentum tensor of a bulk viscous fluid by a term that represents 
a cosmological constant varying with time, and later generalize the solutions of Refs. 
\cite{ref35,ref40,ref41}. This paper is organized as follows: The field equations and the 
generation technique are presented in Section $2$. We relate three of the metric variables 
by solving the off-diagonal component of EFEs, and find a second integral which is used 
to relate the remaining two metric variables. In Section 3, for the particular form of 
each metric variables, some solutions are presented separately and solutions of 
Camci {\it et al.} \cite{ref40}, Ram \cite{ref35} and Pradhan {\it et al.} \cite{ref41} are 
shown to be particular cases of our solutions. Kinematical and dynamical properties of 
all solutions are also studied in this section. In Section $4$, we give the concluding remarks. 

%%%%%%%%%%%%%%%%%%%%%%%%%%%%%%%%%%%%%% SECTION 2 %%%%%%%%%%%%%%%%%%%%%%%%%%%%%%%%%%%%%%%
\section{Field equations and generation technique} 
In this section, we review the solutions obtained by Pradhan {\it et al} \cite{ref41} .
The usual energy-momentum tensor is modified by addition of a term
\begin{equation}
\label{eq1}
T^{(vac)}_{ij} = - \Lambda(t) g_{ij},
\end{equation}
where $\Lambda(t)$ is the cosmological term and $g_{ij}$ is the metric tensor.
Thus the new stress energy-momentum tensor in presence of bulk stress is given by
\begin{equation}
\label{eq2}
T_{ij} = (\bar{p} + \rho)u_{i}u_{j} - \bar{p} g_{ij} - \Lambda(t) g_{ij},
\end{equation}
where 
\begin{equation}
\label{eq3}
\bar{p} = p + \xi u^{i}_{;i}
\end{equation}
Here, $\rho$, $p$, $\bar{p}$, $\xi$ and $u$ are, respectively, the energy density, isotropic 
pressure, effective pressure, bulk viscous coefficient and the fluid four-velocity vector of 
distribution such that $u^{i} u_{i} = 1$. In general, $\xi$ is a function of time.

We consider the space-time metric of the spatially homogeneous Bianchi type V of
the form
\begin{equation}
\label{eq4}
ds^{2} = dt^{2} - A^{2}(t) dx^{2} - e^{2\alpha x} \left[B^{2}(t) dy^{2} + C^{2}(t)dz^{2}\right],
\end{equation}
where $\alpha$ is a constant. For the energy momentum tensor (\ref{eq2}) and Bianchi 
type V space-time (\ref{eq4}), Einstein's field equations
\begin{equation}
\label{eq5}
R_{ij} - \frac{1}{2} R g_{ij} = - 8\pi T_{ij}
\end{equation}
yield the following five independent equations
\begin{equation}
\label{eq6}
\frac{A_{44}}{A} + \frac{B_{44}}{B} + \frac{A_{4}B_{4}}{AB} - \frac{\alpha^{2}}
{A^{2}} = - 8\pi (\bar{p} + \Lambda),
\end{equation}
\begin{equation}
\label{eq7}
\frac{A_{44}}{A} + \frac{C_{44}}{C} + \frac{A_{4}C_{4}}{AC} - \frac{\alpha^{2}}
{A^{2}} = - 8\pi (\bar{p} + \Lambda),
\end{equation}
\begin{equation}
\label{eq8}
\frac{B_{44}}{B} + \frac{C_{44}}{C} + \frac{B_{4}C_{4}}{BC} - \frac{\alpha^{2}}
{A^{2}} = - 8\pi (\bar{p} + \Lambda),
\end{equation}
\begin{equation}
\label{eq9}
\frac{A_{4}B_{4}}{AB} + \frac{A_{4}C_{4}}{AC} + \frac{B_{4}C_{4}}{BC} - 
\frac{3\alpha^{2}}{A^{2}} = 8\pi (\rho - \Lambda),
\end{equation}
\begin{equation}
\label{eq10}
\frac{2A_{4}}{A} - \frac{B_{4}}{B} - \frac{C_{4}}{C} = 0.
\end{equation}
Here and in what follows the suffix $4$ by the symbols $A$, $B$, $C$ and $\rho$ denote 
differentiation with respect to $t$. The Bianchi identity ($T^{ij}_{;j} = 0$) takes the 
form
\begin{equation}
\label{eq11}
{\rho}_{4} + (\rho + p) \theta = 0.
\end{equation}
It is worth noting here that our approach suffers from a lack of Lagrangian 
approach. There is no known way to present a consistent Lagrangian model
satisfying the necessary conditions discussed in this paper. 

The physical quantities expansion scalar $\theta$ and shear scalar $\sigma^{2}$
have the following expressions:
\begin{equation}
\label{eq12}
\theta = u^{i}_{;i} = \frac{A_{4}}{A} + \frac{B_{4}}{B} + \frac{C_{4}}{C}
\end{equation} 
\begin{equation}
\label{eq13}
\sigma^{2} = \frac{1}{2}\sigma_{ij}\sigma^{ij} = \frac{1}{3}\left[\theta^{2} - 
\frac{A_{4}B_{4}}{AB} - \frac{A_{4}C_{4}}{AC} - \frac{B_{4}C_{4}}{BC}\right].
\end{equation} 
Integrating Eq. (\ref{eq10}) and absorbing the integration constant into $B$ or
$C$, we obtain
\begin{equation}
\label{eq14}
A^{2} = BC
\end{equation} 
without any loss of generality. Thus, elimination of $\bar{p}$ from
Eqs. (\ref{eq6}) - (\ref{eq8}) gives the condition of isotropy of pressures
\begin{equation}
\label{eq15}
2\frac{B_{44}}{B} + \left(\frac{B_{4}}{B}\right)^{2} = 2\frac{C_{44}}{C} + 
\left(\frac{C_{4}}{C}\right)^{2},
\end{equation} 
which on integration yields
\begin{equation}
\label{eq16}
\frac{B_{4}}{B} - \frac{C_{4}}{C} = \frac{k}{(BC)^{3/2}},
\end{equation}
where $k$ is a constant of integration. Hence for the metric function $B$ or 
$C$ from the above first order differential Eq. (\ref{eq16}), some scale 
transformations permit us to obtain new metric function $B$ or $C$. \\

Firstly, under the scale transformation $dt = B^{1/2}d\tau$, Eq. (\ref{eq16}) 
takes the form     
\begin{equation}
\label{eq17}
C B_{\tau} - B C_{\tau} = k C^{-1/2},
\end{equation}
where subscript represents derivative with respect to $\tau$. Considering Eq. 
(\ref{eq17}) as a linear differential equation for $B$, where $C$ is an arbitrary
function, we obtain
\begin{equation}
\label{eq18}
(i) B = k_{1} C + k C \int{\frac{d\tau}{C^{5/2}}},
\end{equation}
where $k_{1}$ is an integrating constant. Similarly, using the transformations
$dt = B^{3/2}d\tilde{\tau}$, $dt = C^{1/2}dT$, and $dt = C^{3/2} d\tilde{T}$
in Eq. (\ref{eq16}), after some algebra we obtain respectively 
\begin{equation}
\label{eq19}
(ii) B(\tilde{\tau}; k_{2}, k) = k_{2} C ~ \exp{\left(k\int{\frac{d\tilde{\tau}}
{C^{3/2}}}\right)}, 
\end{equation}
\begin{equation}
\label{eq20}
(iii) C(T; k_{3}, k) = k_{3} B - k B \int{\frac{dT}{B^{5/2}}},
\end{equation}
and
\begin{equation}
\label{eq21}
(iv) C(\tilde{T}; k_{4}, k) = k_{4} B ~ \exp{\left(k\int{\frac{d\tilde{T}}
{B^{3/2}}}\right)}, 
\end{equation}
where $k_{2}$, $k_{3}$ and $k_{4}$ are constants of integration. Thus choosing 
any given function $B$ or $C$ in cases (i), (ii), (iii) and (iv), one can obtain 
$B$ or $C$ and hence $A$ from (\ref{eq14}).

%%%%%%%%%%%%%%%%%%%%%  SECTION 3  %%%%%%%%%%%%%%%%%%%%%%%%%%%%%%%%%%%%%%%%%%
\section{Generation of new  solutions} 

We consider the following four cases:  

%%%%%%%%%%%%%%%%%%%%%%%%%%%%%%%%% SUBSECTION 3.1  %%%%%%%%%%%%%%%%%%%%%%%%%%%%%%%%%
\subsection{Case (I): Let $C = \tau^{n}$, ($n$ is a real number satisfying
$n \ne \frac{2}{5}$).}  
In this case, Eq. (\ref{eq18}) gives
\begin{equation}
\label{eq22}
B = k_{1}\tau^{n} + \frac{2k}{2 - 5n}\tau^{1 - 3n/2}
\end{equation}
and then from (\ref{eq14}), we obtain
\begin{equation}
\label{eq23}
A^{2} = k_{1}\tau^{2n} + \frac{2k}{2-5n}\tau^{1 - n/2}.
\end{equation}
Hence the metric (\ref{eq4}) reduces to the new form
\begin{equation}
\label{eq24}
ds^{2} = \left(k_{1}\tau^{n} + 2\ell \tau^{\ell_{1}}\right)[d\tau^{2} - \tau^{n} 
dx^{2}] - e^{2\alpha x}\left[\left(k_{1}\tau^{n} + 2\ell \tau^{\ell_{1}}\right)^
{2} dy^{2} + \tau^{2n} dz^{2}\right],
\end{equation}
where
\[
\ell = \frac{k}{2 - 5n} ~ and ~ \ell_{1} = 1 - \frac{3n}{2}.
\]
For this derived model (\ref{eq24}), the effective pressure, energy density and cosmological 
constant are given by
\[
8\pi(\bar{p} + \Lambda) = \left(k_{1}\tau^{n} + 2\ell \tau^{\ell_{1}}\right)^{-3}
\biggl[-2k^{2}_{1} n(n - 1)\tau^{2n -2} - k_{1}\ell n(10 - 13n)\tau^{-(\ell_{1} + 2n)}
\] 
\begin{equation}
\label{eq25}
- \frac{\ell^{2}(4 + 4n - 11n^{2})}{2}\tau^{-3n}\biggr] + \alpha^{2}\tau^{-n}
\left(k_{1} \tau^{n} + 2\ell \tau^{\ell_{1}}\right)^{-1},
\end{equation}
\[
8\pi(\rho - \Lambda) = \left(k_{1}\tau^{n} + 2\ell \tau^{\ell_{1}}\right)^{-3}
\biggl[3k^{2}_{1} n^{2}\tau^{2n -2} + 3k_{1}\ell n(2 - n)\tau^{-(\ell_{1} + 2n)}
\] 
\begin{equation}
\label{eq26}
+ \frac{\ell^{2}(4 + 4n - 11n^{2})}{2}\tau^{-3n}\biggr] - 3\alpha^{2}\tau^{-n}
\left(k_{1} \tau^{n} + 2\ell \tau^{\ell_{1}}\right)^{-1}.
\end{equation}
We assume for the specification of $\xi$, that the fluid obeys an equation of state of the form
\begin{equation}
\label{eq27} 
p = \gamma \rho,
\end{equation}
where $\gamma(0 \leq \gamma \leq 1)$ is a constant. 

Thus, given $\xi(t)$, we can solve the system for the physical
quantities. In most of investigations involving bulk viscosity is assumed 
to be a simple power function of the energy density (Pavon {\it et al.}\cite{ref42}; 
Maartens \cite{ref43}; Zimdahl \cite{ref44}; Santos {\it et al.} \cite{ref45})
\begin{equation}
\label{eq28} 
\xi(t) = \xi_{0} \rho^{w},
\end{equation}
where $\xi_{0}$ and $w$ are real constants. For small density, $w$ may even be 
equal to unity as used in Murphy's work \cite{ref46} for simplicity. If $w = 1$, Eq.
(\ref{eq28}) may correspond to a radiative fluid (Weinberge \cite{ref14}). Near the 
big bang, $0 \leq w \leq \frac{1}{2}$ is a more appropriate assumption (Belinskii 
and Khalatnikov \cite{ref47}) to obtain realistic models.

For simplicity and realistic models of physical importance, we consider the following 
two cases ($w = 0, 1$):  

%%%%%%%%%%%%%%%%%%%%%%%%%%%%%%%%%%%%%%%%%%%%%%%%%%%%%%%%%%%%%%%%%%%%%%%%%%%%%%%%%%%%%%
%%%%%%%%%%%%%%%%%%%%%%%%%%%%%%%%% SUBSUBSECTION 3.1.1  %%%%%%%%%%%%%%%%%%%%%%%%%%%%%%%%%
\subsubsection{Model I: ~ ~ Solution for $w = 0$}
When $w = 0$, Eq. (\ref{eq28}) reduces to $\xi = \xi_{0}$ = constant. Hence in this case 
Eq. (\ref{eq25}), with the use of (\ref{eq26}), (\ref{eq27}) and (\ref{eq28}), leads to
\[
8\pi(1 + \gamma)\rho = {D_{1}}^{-3}\Big[n(n + 2)k_{1}^{2}\tau^{2(n - 1)} + 2(5n -2)k_{1} \ell n 
\tau^{-(\ell_{1} + 2n)}\Big] +
\]
\begin{equation}
\label{eq29}
 {D_{1}}^{-1}\Big[-2\alpha^{2}\tau^{-n} + 24\pi\xi_{0}\left(k_{1} n \tau^{n -1}
+ \frac{1}{2}\ell(2 - n)\tau^{-\frac{3n}{2}}\right){D_{1}}^{-\frac{1}{2}}\Big], 
\end{equation}  
where 
$$D_{1} = k_{1}\tau^{n} + 2\ell \tau^{\ell_{1}}.$$

Eliminating $\rho(t)$ from Eqs. (\ref{eq26}) and (\ref{eq29}), we obtain  
\[
8\pi (1 + \gamma)\Lambda = {D_{1}}^{-3}\Big[(1 - 2n - 3n\gamma)n k^{2}_{1}\tau^{2(n - 1)} - 
\{(10 - 13n) + 3(2 - n)\gamma\}k_{1}\ell n \tau^{-(\ell_{1} + 2n)}
\]
\[
- \frac{1}{2}(4 + 4n - 11n^{2})(1 + \gamma)\ell^{2}\tau^{-3n}\Big] +
\]
\begin{equation}
\label{eq30}
(1 + 3\gamma)\alpha^{2}\tau^{-n}{D_{1}}^{-1} + 24\pi\xi_{0}\left[nk_{1}
\tau^{n - 1} + \frac{1}{2}\ell(2 - n)\tau^{-\frac{3n}{2}}\right]{D_{1}}^{-\frac{3}{2}}
\end{equation}

%%%%%%%%%%%%%%%%%%%%%%%%%%%%%%%%%%%%%%%%%%%%%%%%%%%%%%%%%%%%%%%%%%%%%%%%%%%%%%%%%%%%%%
%%%%%%%%%%%%%%%%%%%%%%%%%%%%%%%%% SUBSUBSECTION 3.1.2  %%%%%%%%%%%%%%%%%%%%%%%%%%%%%%%%%
\subsubsection{Model II: ~ ~ Solution for $w = 1$}
When $w = 1$, Eq. (\ref{eq28}) reduces to $\xi = \xi_{0}\rho$. Hence in this case 
Eq. (\ref{eq25}), with the use of (\ref{eq26}), (\ref{eq27}) and (\ref{eq28}), leads to
\begin{equation}
\label{eq31}
8\pi \rho = \frac{\Big[n(n + 2)k^{2}_{1}\tau^{2(n - 1)} + 2n(5n - 2)k_{1}\ell \tau^{-(\ell_{1} + 2n)} 
- 2\alpha^{2}\tau^{-n}{D_{1}}^{2}\Big]}{{D_{1}}^{3}\left[(1 + \gamma) - 3\xi_{0}\{nk_{1}\tau^{n -1} + 
\frac{1}{2}(2 - n)\ell \tau^{-\frac{3n}{2}}\}{D_{1}}^{-\frac{3}{2}}\right]}. 
\end{equation}
Eliminating $\rho(t)$ between (\ref{eq26}) and (\ref{eq31}), we obtain 
\[
8\pi \Lambda = \frac{\Big[n(n + 2)k^{2}_{1}\tau^{2(n - 1)} + 2n(5n - 2)k_{1}\ell \tau^{-(\ell_{1} + 2n)} 
- 2\alpha^{2}\tau^{-n}{D_{1}}^{2}\Big]}{{D_{1}}^{3}\left[(1 + \gamma) - 3\xi_{0}\{nk_{1}\tau^{n -1} + 
\frac{1}{2}(2 - n)\ell \tau^{-\frac{3n}{2}}\}{D_{1}}^{-\frac{3}{2}}\right]} 
\]
\begin{equation}
\label{eq32}
- \frac{1}{{D_{1}}^{3}}\left[3n^{2}k^{2}_{1}\tau^{2(n - 1)} + 3n (2 - n)k_{1}\ell \tau^{-(\ell_{1} + 2n)} 
+\frac{1}{2}(4 + 4n - 11n^{2})\ell^{2}\tau^{-3n} - 3\alpha^{2}\tau^{-n} {D_{1}}^{2}\right].
\end{equation}

From Eqs. (\ref{eq29}) and (\ref{eq30}), we observe that at the time of early universe the 
the energy density $\rho(t)$ and cosmological constant ($\Lambda(t)$) decrease with time increases
(see Figure 1). We also observe that the value of $\Lambda$ is small and positive at late times 
which is supported by recent type Ia supernovae observations \cite{ref9} $-$ \cite{ref12}. In Model II, 
from Eqs. (\ref{eq31}) and (\ref{eq32}), we find that for large range of parameters the energy density 
decreases with time very sharply and becomes negative (even if it is positive initially) and then remains 
negative throughout the evolution. The cosmological constant shows singular behaviour closer to origin. 
Then it shows eretic behaviour and later stage it remain negative constant value. It seems that Model II 
may not be physical model of the universe. \\  

The metric (\ref{eq24}) is a four-parameter family of solutions to EFEs with a
bulk viscous fluid. Using the scale transformation $dt = B^{\frac{1}{2}} d\tau$ in Eqs. (\ref{eq12})
and (\ref{eq13}) for this case, the scalar expansion $\theta$ and the shear $\sigma$ 
have the expressions:
\begin{equation}
\label{eq33}
\theta = 3\left[k_{1} n \tau^{n - 1} + \frac{\ell(2 - n)}{2} \tau^{-3n/2}\right]
\left(k_{1}\tau^{n} + 2\ell \tau^{\ell_{1}}\right)^{-3/2}
\end{equation} 
\begin{equation}
\label{eq34}
\sigma = \frac{1}{2} k \tau^{-3n/2}\left(k_{1}\tau^{n} + 2\ell \tau^{\ell_{1}}
\right)^{-3/2}
\end{equation}
Eqs. (\ref{eq33}) and (\ref{eq34}) lead to
\begin{equation}
\label{eq35}
\frac{\sigma}{\theta} = \frac{k}{6}\left[k_{1} n \tau^{n - \ell_{1}} + \frac{\ell(2 - n)}
{2}\right]^{-1}
\end{equation}
Now, we consider four subcases for the parameters $\Lambda$, $n$, $k$, $k_{1}$ whether 
zero or not. \\

In subcase $\Lambda = 0$ and $\xi_{0} = 0$, metric (\ref{eq24}) with expressions $p$, $\rho$, $\theta$ 
and $\sigma$ for this model are same as that of solution (\ref{eq18}) of Camci 
{\it et al.} \cite{ref40}. If we set $\xi_{0} = 0$, the metric (\ref{eq24}) gives the solution obtained 
by Pradhan {\it et al.} \cite{ref41}. \\

In subcase $\Lambda = 0$, $n = 0$, after a suitable inverse time transformation, 
we find that 
\begin{equation}
\label{eq36}
ds^{2} = dt^{2} - K_{1}(t + t_{0})^{2/3}dx^{2} - e^{2\alpha x}\left[K_{1}(t + t_{0})^
{4/3}dy^{2} + dz^{2}\right],
\end{equation}
where $t_{0}$ is a constant of integration and $K_{1} = (3k/2)^{2/3}$. The expressions 
$p$, $\rho$, $\theta$ and $\sigma$ for this model are not given here, since it is 
observed that the physical properties of this one are same as that of the solution 
(\ref{eq24}) of Ram \cite{ref35}. \\

In subcase $\Lambda = 0$, $k =0$, after inverse time transformation and rescaling,
the metric (\ref{eq24}) reduces to
\begin{equation}
\label{eq37}
ds^{2} = dt^{2} - K_{2}(t + t_{1})^{\frac{4n}{n + 2}}\left[dx^{2} + e^{2\alpha x}
(dy^{2} + dz^{2})\right],
\end{equation}  
where $t_{1}$ is a constant of integration and $K_{2} = \left(\frac{n + 2}{2}\right)
^{\frac{4n}{n + 2}}$. For this solution, when $n = 1$ and $\alpha = 0$, we obtain
Einstein and de Sitter [48] dust filled universe. For $K_{2} = 1$, $t_{1} 
= 0$ and $ n = \frac{2m}{(2 - m)}$, where $m$ is a parameter in Ram's paper [35], 
the solution (\ref{eq37}) reduces to the metric (\ref{eq14}) of Ram \cite{ref35}. In later case, 
if also $\alpha = 0$, then we get the Minkowski 
space-time. \\

Now, in subcase $\Lambda = 0$, $k_{1} = 0$, after some algebra the metric (\ref{eq24})
takes the form
\begin{equation}
\label{eq38}
ds^{2} = dt^{2} - 2\ell K_{3} (t + t_{2})^{2/3}\left[dx^{2} + e^{2\alpha x} \left(
a t^{m_{1}} dy^{2} + a^{-1}t^{-m_{1}} dz^{2}\right)\right],
\end{equation} 
where $t_{2}$ is a constant, $at^{m_{1}} = 2\ell K^{\frac{2 - 5n}{2 - n}}_{3} (t + t_{2})
^{\frac{2(2 - 5n)}{3(2 - n)}}$ and $ K_{3} = \left[\frac{3(2 - n)}{4\sqrt{2 \ell}}
\right]^{2/3}$. \\
For $t_{2} = 0$, $k = \frac{2}{3}$ and $n = 0$ from (\ref{eq38}), we obtain that the 
solution (\ref{eq24}) of Ram \cite{ref35}.\\  

%%%%%%%%%%%%%%%%%%%%%%%%%%%%%%%%%%%%%%%%%%%%%%%%%%%%%%%%%%%%%%%%%%%%%%%%%%%%%%%%%%%%%%%%%

{\bf {Some Physical aspects of model} :} \\

The model (\ref{eq24}) has barrel singularity at $\tau = \tau_{0}$ given by
\[
\tau_{0} = \left[\frac{k_{1}(5n - 2)}{2k}\right]^{\frac{2}{(2 - 5n)}},
\]
which corresponds to $t = 0$. For $n \ne 2/5$ from (\ref{eq24}), it is observed that 
at the singularity state $\tau = \tau_{0}$, $p$, $\rho$, $\Lambda$, $\theta$ and 
$\sigma$ are infinitely large. At $t \to \infty$, which corresponds to $\tau \to 
\infty$ for $n < 2/5$ and $k> 0$, or $\tau \to 0$ for $n > 2/5$ and $k < 0$, $p$, $\rho$, 
$\Lambda$, $\theta$ and $\sigma$ vanish. Therefore, for $n \ne 2/5$, the solution
(\ref{eq24}) represents an anisotropic universe exploding from $\tau = \tau_{0}$, i.e.
$t = 0$, which expands for $0 < t < \infty$. We also find that the ratio $\sigma / \theta$
tends to a finite limit as $t \to \infty$, which means that the shear scalar does not 
tend to zero faster than the expansion. Hence the model does not approach isotropy for
large values of $t$. 

In subcase $\Lambda = 0$, $k = 0$, the ratio (\ref{eq35}) tends to zero, then the model
approaches isotropy i.e. shear scalar $\sigma$ goes to zero. For the model (\ref{eq37}),
$p$ and $\rho$ tends to zero as $t \to \infty$; the model would give an essentially empty
universe at large time. The dominant energy condition given by Hawking and Ellis \cite{ref49}  
requires that 
\begin{equation}
\label{eq39}
\rho + p \geq 0, ~ ~ \rho + 3p \geq 0
\end{equation} 
Thus, we find for the model (\ref{eq37}) that $n(2 - n) \geq 0$. Hence for the values
$0 \leq n \leq 2$, the universe (\ref{eq37}) satisfies the strong energy condition i.e.
$\rho + 3p \geq 0$. Also this model is sheer-free and expanding. \\ 

In subcase $\Lambda = 0$, $k_{1} = 0$, for $n \neq 2/5$, $2$, it is observed from relations
(\ref{eq29}) - (\ref{eq34}) that $p$, $\rho$, $\theta$ and $\sigma$ are infinitely large
at the singularity state $t = - t_{2}$. When $t \to \infty$, these quantities vanish.
We also find that the ratio $\sigma/ \theta$ is a constant. This shows that the cosmological
 model (\ref{eq38}) does not approach isotropy for large value of $t$. In this model the
dominant energy conditions (\ref{eq39}) are then verified for $6 - 5n - 25n^{2} \geq 0$.
Since $n \ne 2/5$, the model (\ref{eq38}) satisfies the strong energy condition for 
$-3/5 \leq n \leq 2/5$. \\

In each of subcases, all the obtained solutions (\ref{eq36}), (\ref{eq37}) and 
(\ref{eq38}) satisfy the Bianchi identity given in Eq. (\ref{eq10}).  
%%%%%%%%%%%%%%%%%%%%%%% Figure 1 %%%%%%%%
\begin{figure}[ht] %ORIGINAL SIZE: width=1.4TRUEIN; height=1.5TRUEIN
\vspace*{13pt}
\centerline{\includegraphics[width=1.00\textwidth,angle=0]{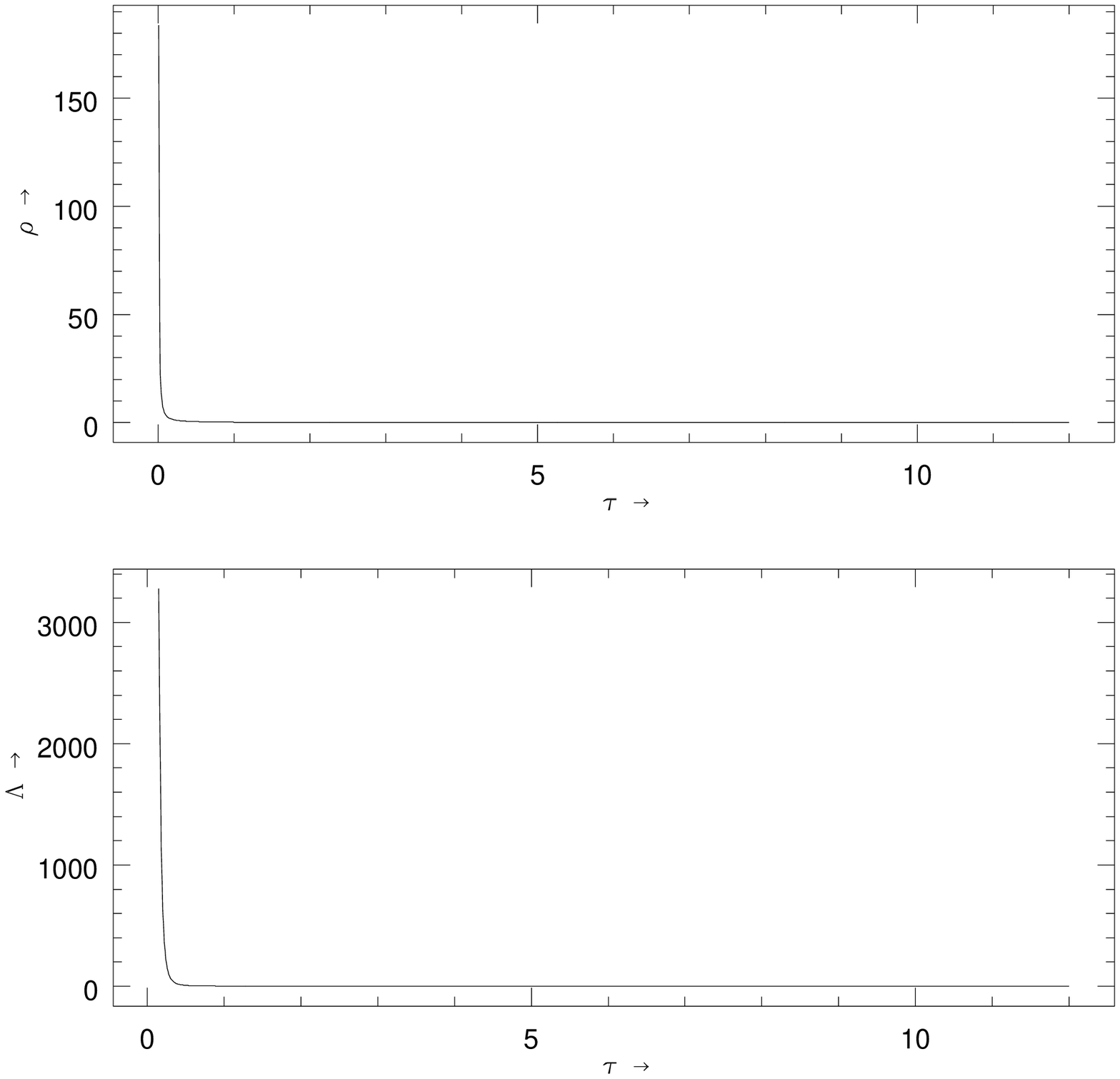}}
%%\centerline{\psfig{file=fig2.eps,width=8cm}} %100 percent
\vspace*{13pt}
\caption{(Top) Plot of $\rho \rightarrow \tau$ and (bottom) $\Lambda \rightarrow
\tau$ for parameters $n=1/4$, $k_1 = 1$, $\gamma = 0.5$, $\alpha=1$, $\xi_{0} = 1$ and
rest of the constants are set to 1.}
\label{fig1}
\end{figure}
%%%%%%%%%%%%%%%%%%%%%%%%%%%%%%%%%%%%%%%%%%%%%%%%%%%%%%%%%%%%%%%%%%%%%%%%%%%%%%%%
%%%%%%%%%%%%%%%%%%%%%%%%%%%%%%%%%  SUBSECTION 3.2 %%%%%%%%%%%%%%%%%%%%%%%%%%%%%%

\subsection{Case (II): Let $C = \tilde{\tau}^{n}$, ($n$ is a real number 
satisfying $n \ne 2/3$).} 
In this case Eq. (\ref{eq19}) gives
\begin{equation}
\label{eq40}
B = k_{2} \tilde{\tau}^{n} \exp{\left(M\tilde{\tau}^{\ell_{1}}\right)}
\end{equation} 
and from (\ref{eq14}), we obtain
\begin{equation}
\label{eq41}
A^{2} = k_{2} \tilde{\tau}^{2n} \exp{\left(M\tilde{\tau}^{\ell_{1}}\right)}
\end{equation} 
where $M = \frac{k}{\ell_{1}}$. Hence the metric (\ref{eq4}) reduces to the form
\[
ds^{2} = \tilde{\tau}^{4(1 - \ell_{1})/3}\biggl[\tilde{\tau}^{2(1 - \ell_{1})/3} 
e^{3M\tilde{\tau}^{\ell_{1}}} d\tilde{\tau}^{2} - e^{M\tilde{\tau}^{\ell_{1}}} dx^{2}
\]  
\begin{equation}
\label{eq42}
- e^{2\alpha x}\left(e^{2M\tilde{\tau}^{\ell_{1}}} dy^{2} + dz^{2}\right)\biggr],
\end{equation}
where the constant $k_{2}$ is taken, without any loss of generality, equal to $1$.
This metric is a three-parameter family of solutions to EFEs with a perfect fluid.\\

For the above model, the distribution of matter and nonzero kinematical parameters
are obtained as  
\[
8\pi(\bar{p} + \Lambda) = 2n\tilde{\tau}^{2(\ell_{1} - 2)} + 3nk\tilde{\tau}^{3\ell_{1} - 4}
\]
\begin{equation}
\label{eq43}
+ \frac{k^{2}}{2}\tilde{\tau}^{4(\ell_{1} - 1)} + \alpha^{2}\tilde{\tau}^
{4(\ell_{1} - 1)/3} e^{-3M\tilde{\tau}^{\ell_{1}}},
\end{equation} 
\[
8\pi(\rho - \Lambda) = 3n^{2}\tilde{\tau}^{2(\ell_{1} - 2)} + 3nk\tilde{\tau}^
{3\ell_{1} - 4}
\]
\begin{equation}
\label{eq44}
+ \frac{k^{2}}{2}\tilde{\tau}^{4(\ell_{1} - 1)} - 3\alpha^{2}\tilde{\tau}^
{4(\ell_{1} - 1)/3} e^{-3M\tilde{\tau}^{\ell_{1}}}.
\end{equation} 

For simplicity and realistic models of physical importance, we consider the following 
two cases ($w = 0, 1$):  

%%%%%%%%%%%%%%%%%%%%%%%%%%%%%%%%%%%%%%%%%%%%%%%%%%%%%%%%%%%%%%%%%%%%%%%%%%%%%%%%%%%%%%
%%%%%%%%%%%%%%%%%%%%%%%%%%%%%%%%% SUBSUBSECTION 3.2.1  %%%%%%%%%%%%%%%%%%%%%%%%%%%%%%%%%
\subsubsection{Model I: ~ ~ Solution for $w = 0$}
When $w = 0$, Eq. (\ref{eq28}) reduces to $\xi = \xi_{0}$ = constant. Hence in this case 
Eq. (\ref{eq43}), with the use of (\ref{eq44}), (\ref{eq27}) and (\ref{eq28}), leads to
\begin{equation}
\label{eq45}
8\pi (1 + \gamma)\rho = n(2 + 3n)\tilde{\tau}^{2(\ell_{1} - 2)} + 6nk\tilde{\tau}^{(3\ell_{1} - 4)} 
+ k^{2}\tilde{\tau}^{4(\ell_{1} - 1)} - 2 D_{3} + 24\pi \xi_{0} D_{2},
\end{equation}
where
$$D_{2} = n \tilde{\tau}^{(\ell_{1} - 2)} + \frac{1}{2}k \tilde{\tau}^{2(\ell_{1} - 1)},$$

$$D_{3} = \alpha^{2} \tilde{\tau}^{\frac{4}{3}(\ell_{1} - 1)}e^{-3M \tilde{\tau}^{\ell_{1}}}.$$ 

Eliminating $\rho(t)$ from Eqs. (\ref{eq44}) and (\ref{eq45}), we obtain  
\[
8\pi(1 + \gamma)\Lambda = n(2 - 3n\gamma)\tilde{\tau}^{2(\ell_{1} - 2)} + 3nk(1 - \gamma)\tilde{\tau}^
{(3\ell_{1} - 4)}
\]
\begin{equation}
\label{eq46} 
+ \frac{1}{2}k^{2}(1 - \gamma)\tilde{\tau}^{4(\ell_{1} - 1)} + (1 - 3\gamma)D_{3} + 24\pi \xi_{0} D_{2}. 
\end{equation} 

%%%%%%%%%%%%%%%%%%%%%%%%%%%%%%%%%%%%%%%%%%%%%%%%%%%%%%%%%%%%%%%%%%%%%%%%%%%%%%%%%%%%%%
%%%%%%%%%%%%%%%%%%%%%%%%%%%%%%%%% SUBSUBSECTION 3.2.2  %%%%%%%%%%%%%%%%%%%%%%%%%%%%%%%%%
\subsubsection{Model II: ~ ~ Solution for $w = 1$}
When $w = 1$, Eq. (\ref{eq28}) reduces to $\xi = \xi_{0}\rho$ . Hence in this case 
Eq. (\ref{eq43}), with the use of (\ref{eq44}), (\ref{eq27}) and (\ref{eq28}), leads to
\begin{equation}
\label{eq47} 
8\pi \rho = \frac{n(2 + 3n)\tilde{\tau}^{2(\ell_{1} - 2)} + 6nk \tilde{\tau}^{(3\ell_{1} - 4)} 
+ k^{2}\tilde{\tau}^{4(\ell_{1} - 1)} - 2D_{3}}{(1 + \gamma) - 3\xi_{0}D_{2}}.
\end{equation} 
Eliminating $\rho(t)$ from Eqs. (\ref{eq44}) and (\ref{eq47}), we obtain  
\[
8\pi \Lambda = \frac{n(2 + 3n)\tilde{\tau}^{2(\ell_{1} - 2)} + 6nk \tilde{\tau}^{(3\ell_{1} - 4)} 
+ k^{2}\tilde{\tau}^{4(\ell_{1} - 1)} - 2D_{3}}{(1 + \gamma) - 3\xi_{0}D_{2}}
\]
\begin{equation}
\label{eq48} 
- 3n^(2)\tilde{\tau}^{2(\ell_{1} - 2)} - 3nk \tilde{\tau}^{(3\ell_{1} - 4)} - \frac{1}{2}k^{2} 
\tilde{\tau}^{4(\ell_{1} - 1)} + 3D_{3}.
\end{equation} 

In Mode I, from Eqs. (\ref{eq45}) and (\ref{eq46}), we observe that the energy density $\rho(t)$ 
and cosmological constant ($\Lambda(t)$) decrease with time increases (see Figure 2). Here we find 
energy density always positive. We also observe that the value of $\Lambda$ is small and positive 
at late times which is supported by recent type Ia supernovae observations \cite{ref9} $-$ \cite{ref12}. 
From Eqs. (\ref{eq47}) and (\ref{eq48}), we observe that Model II has the similar behaviour as Model I, 
so it is not reproduced here. \\

The scalar of expansion $\theta$ and the shear $\sigma$ are obtained as
\begin{equation}
\label{eq49}
\theta = 3\left[n\tilde{\tau}^{\ell_{1} - 2} + \frac{k}{2}\tilde{\tau}^{2(\ell_{1} 
- 1)}\right],
\end{equation} 
\begin{equation}
\label{eq50}
\sigma = \frac{k}{2}\tilde{\tau}^{2(\ell_{1} - 1)}e^{-3M\tilde{\tau}^{\ell_{1}}},
\end{equation} 
From Eqs. (\ref{eq49}) and (\ref{eq50}), we have
\begin{equation}
\label{eq51}
\frac{\sigma}{\theta} = \frac{k}{6\left(n \tilde{\tau}^{-\ell_{1}} + \frac{k}{2}\right)}.
\end{equation} 

In subcase $\Lambda = 0$ and $\xi_{0} = 0$, metric (\ref{eq42}) with expressions $p$, $\rho$, $\theta$ 
and $\sigma$ for this model are same as that of solution (\ref{eq27}) of Camci 
{\it et al.} \cite{ref40}. 

In sub-case $\Lambda = 0$, $\xi_{0} = 0$, $\ell_{1} = 1$ (i.e. $n = 0$), we find a similar solution to
(\ref{eq36}), and hence this subclass is omitted. For $k =0$, the ratio (\ref{eq51}) 
is zero and hence there is no anisotropy. \\

After a suitable coordinate transformation, the metric (\ref{eq42}) can be written as
\begin{equation}
\label{eq52}
ds^{2} = dt^{2} - K_{4}(t + t_{3})^{2\ell_{1}}\left[dx^{2} + e^{2\alpha x}(dy^{2} 
+ dz^{2})\right],
\end{equation} 
where $t_{3}$ is a constant and $K_{4} = \left[\frac{2}{(2 - 3M_{1})}\right]^{2M_{1}}$, 
$M_{1} = \frac{2n}{2 + 3n} \ne \frac{2}{3}$, where $M_{1}$ is a new parameter. When $M_{1} = 0$ 
and $\ell_{1} = 0$, from (\ref{eq52}), we get the solution (\ref{eq12}) of Ram \cite{ref35}.\\

\begin{figure}[ht] %ORIGINAL SIZE: width=1.4TRUEIN; height=1.5TRUEIN
\vspace*{13pt}
\centerline{\includegraphics[width=1.00\textwidth,angle=0]{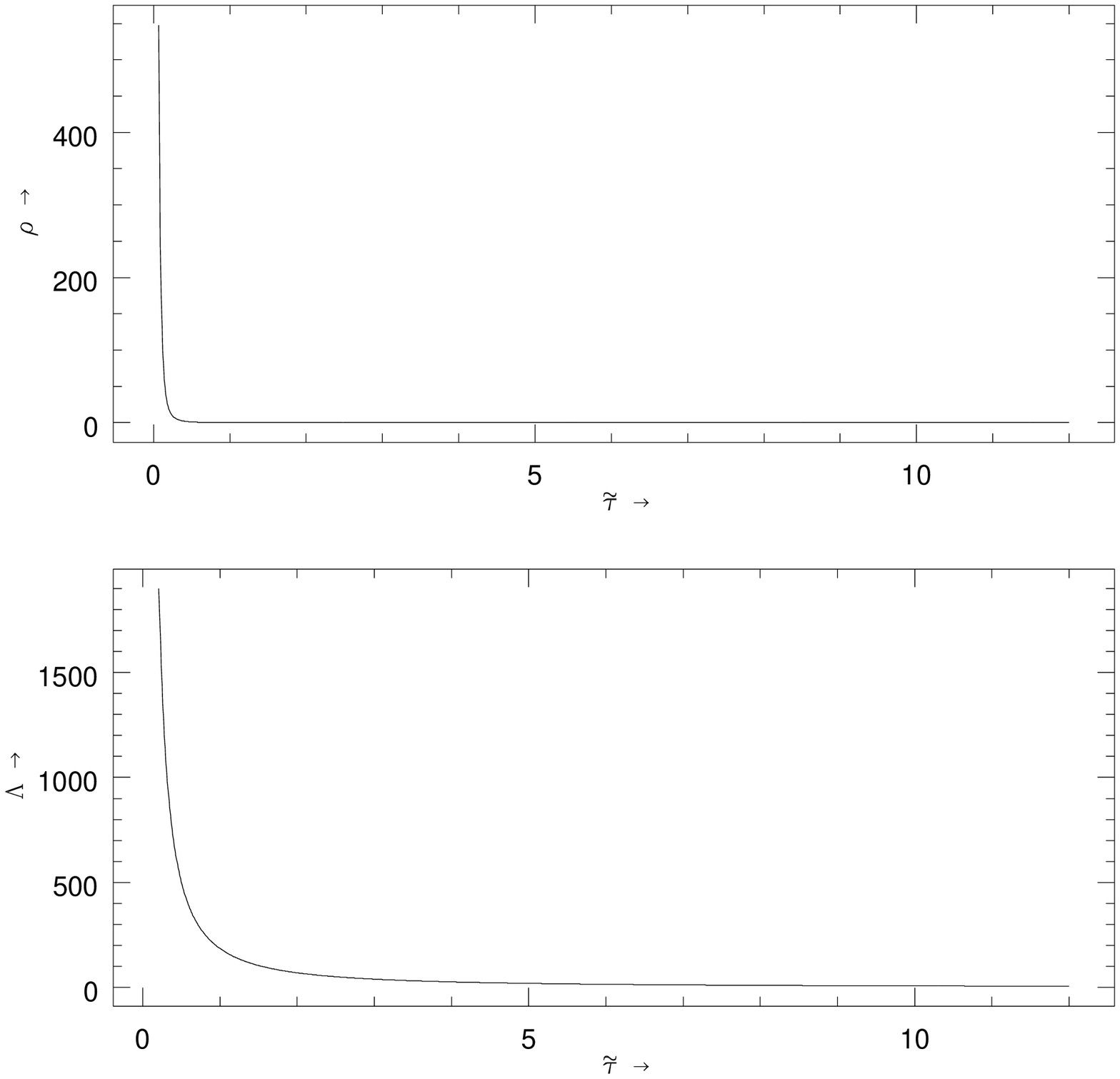}}
%%\centerline{\psfig{file=fig2.eps,width=8cm}} %100 percent
\vspace*{13pt}
\caption{(Top) Plot of $\rho \rightarrow {\tilde{\tau}}$ and (bottom) $\Lambda \rightarrow
{\tilde{\tau}}$ for parameters $n=0.45$, $K=2.0$,  $k_1 = 2$, $\gamma = 0.5$, $\alpha=1$, $\xi_{0} = 1$, 
$M=1$ and rest of the constants are set to 1.}
\label{fig2}
\end{figure}

{\bf {Some physical aspects of model}:} 

The models have singularity at $\tilde{\tau} \to - \infty$ for $\ell_{1} > 0$
or $\tilde{\tau} \to 0$ for $\ell_{1} < 0$, which corresponds to $t \to 0$.
It is a point type singularity for $\ell_{1} > 0$ whereas it is a cigar or a barrel 
singularity according as $\ell_{1} < 0$. At $t \to \infty$, which correspond to
$\tilde{\tau} \to \infty$ for $\ell_{1} > 0$ or $\tilde{\tau} \to 0$ for 
$\ell_{1} < 0$, from Eqs. (\ref{eq45}) - (\ref{eq50}), we obtain that for
$\ell_{1} > 0$, $p,\rho \to 0$, and $\sigma, \theta \to 0$ ($k > 0$), - $\infty$
($k < 0$); for  $\ell_{1} < 0$, similar the above ones. Then, clearly, for a
realistic universe, it must be fulfill as $\tilde{\tau} \to - \infty$, $n$ and $k$ 
are positive and $\ell_{1}$ is an odd positive number; as $\tilde{\tau} \to 0$,
$k$ is positive, and $\ell_{1}$ an even negative number. Also, since $lim_{\tilde{\tau} \to
\infty} \frac{\sigma}{\theta} \ne 0$, therefore these models do not approach isotropy
for large values of $\tilde{\tau}$. \\

In subcase $k = 0$ for the metric (\ref{eq52}), the effective pressure, density and cosmological
constant are given by
\begin{equation}
\label{eq53}
8\pi (\bar{p} + \Lambda) = \frac{\ell_{1}(2 - 3\ell_{1})}{(t + t_{3})^{2}} + \frac{\alpha^{2}}
{K_{4}(t + t_{3})^{2\ell_{1}}},
\end{equation} 
\begin{equation}
\label{eq54}
8\pi (\rho - \Lambda) = \frac{3\ell_{1}}{(t + t_{3})^{2}} - \frac{3\alpha^{2}}
{K_{4}(t + t_{3})^{2\ell_{1}}}.
\end{equation}

When $\Lambda = 0$ and $\xi_{0} = 0$, the pressure and energy density are same as that of given in Eq.
(\ref{eq44}) of paper Camci {\it et al.} \cite{ref40}. In this case, the weak and strong 
energy conditions (\ref{eq39}) for this solution are identically satisfied when 
$\ell_{1}(1 - \ell_{1}) \geq 0$ i.e. $0 \leq \ell_{1} \leq 1$. This model is shear-free 
and expanding with $\theta = \frac{3\ell_{1}}{(t + t_{3})}$.  

%%%%%%%%%%%%%%%%%%%%%%%%%%%%%%%%%%%%%%%%%%%%%%%%%%%%%%%%%%%%%%%%%%%%%%%%%%%%%%%%%%%
%%%%%%%%%%%%%%%%%%%%%%%%%%% SUBSECTION 3.3  %%%%%%%%%%%%%%%%%%%%%%%%%%%%%%%%%%

\subsection{Case (III) : Let $B$ = $T^{n}$ ($n$ is a real number).} 
In this case Eq. (\ref{eq20}) gives
\begin{equation}
\label{eq55}
C = k_{3} T^{n} - 2\ell T^{\ell_{1}}
\end{equation}
and then from (\ref{eq14}), we obtain
\begin{equation}
\label{eq56}
A^{2} = k_{3} T^{2n} - 2\ell T^{\ell_{1} + n}
\end{equation}
Hence the metric (\ref{eq4}) takes the new form
\[
ds^{2} = \left(k_{3} T^{n} - 2\ell T^{\ell_{1}}\right)[dt^{2} - T^{n}dx^{2}] -
\]
\begin{equation}
\label{eq57}
e^{2\alpha x}\left[T^{2n}dy^{2} + \left(k_{3}T^{n} - 2\ell T^{\ell_{1}}\right)^{2}
dz^{2}\right]
\end{equation}
For four-parameters family of solution (\ref{eq57}), the physical and kinematical 
quantities are given by
\[
8\pi(\bar{p} + \Lambda) = \biggl[- \frac{\ell^{2}}{2}(11n^{2} - 4n -4) T^{-3n} + \ell k_{3}
n(13n -10)T^{\ell_{1} + n} - 
\]
\begin{equation}
\label{eq58}
2k^{2}_{3} n(n - 1)T^{2n - 2}\biggr]\left(k_{3}T^{n} - 2\ell T^{\ell_{1}}\right)^{-3} 
+ \alpha^{2} T^{-n}\left(k_{3}T^{n} - 2\ell T^{\ell_{1}}\right)^{-1}, 
\end{equation}
\[
8\pi(\rho - \Lambda) = \biggl[- \frac{\ell^{2}(11n^{2} - 4n - 14)}{2} T^{-3n} - 3\ell k_{3}
n(2 - n)T^{\ell_{1} + n} + 
\]
\begin{equation}
\label{eq59}
3k^{2}_{3} n^{2} T^{2n - 2}\biggr]\left(k_{3}T^{n} - 2\ell T^{\ell_{1}}\right)^{-3} - 
3\alpha^{2} T^{-n}\left(k_{3}T^{n} - 2\ell T^{\ell_{1}}\right)^{-1}. 
\end{equation}
%%%%%%%%%%%%%%%%%%%%%%%%%%%%%%%%%%%%%%%%%%%%%%%%%%%%%%%%%%%%%%%%%%%%%%%%%%%%%%%%%%%%%%
%%%%%%%%%%%%%%%%%%%%%%%%%%%%%%%%% SUBSUBSECTION 3.3.1  %%%%%%%%%%%%%%%%%%%%%%%%%%%%%%%%%
\subsubsection{Model I: ~ ~ Solution for $w = 0$}
When $w = 0$, Eq. (\ref{eq28}) reduces to $\xi = \xi_{0}$ = constant. Hence in this case 
Eq. (\ref{eq58}), with the use of (\ref{eq59}), (\ref{eq27}) and (\ref{eq28}), leads to
\[
8\pi (1 + \gamma)\rho = \left[n(n + 2){k_{3}}^{2}T^{2(n - 1)} + 16n(n - 1)k_{3} \ell T^{(\ell_{1} + n)} 
- (11n^{2} - 4n - 9)\ell^{2} T^{-3n}\right]{D_{4}}^{-3}
\]
\begin{equation}
\label{eq60}
- 2\alpha^{2} T^{-n}{D_{4}}^{-1} + 24 \pi \xi_{0} D_{5} {D_{4}}^{-\frac{3}{2}},
\end{equation}
where

$$D_{4} = k_{3}T^{n} - 2\ell T^{\ell_{1}},$$
$$D_{5} = k_{3} n T^{n -1} + \frac{1}{2}(n - 2)\ell T^{-\frac{3n}{2}}.$$

Eliminating $\rho(t)$ from Eqs. (\ref{eq59}) and (\ref{eq60}), we obtain  
\[
8\pi(1 + \gamma)\Lambda = \Big[\{-\frac{1}{2}n\ell^{2}(1 - \gamma)(11n - 4) - \ell^{2}(7\gamma - 2)\}
T^{-3n} 
\]
\[
+ n \ell k_{3}\{13n - 10 + 3\gamma(2 - n)\}T^{\ell_{1} + n} - n{k_{3}}^{2}(2n - 2 + 3n\gamma)
T^{2(n - 1)}\Big] {D_{4}}^{-3} 
\]
\begin{equation}
\label{eq61}
+ (1 + 3\gamma)\alpha^{2}T^{-n}{D_{4}}^{-n} + 24 \pi \xi_{0} D_{5} {D_{4}}^{-\frac{3}{2}}.
\end{equation}

\begin{figure}[ht] %ORIGINAL SIZE: width=1.4TRUEIN; height=1.5TRUEIN
\vspace*{13pt}
\centerline{\includegraphics[width=1.00\textwidth,angle=0]{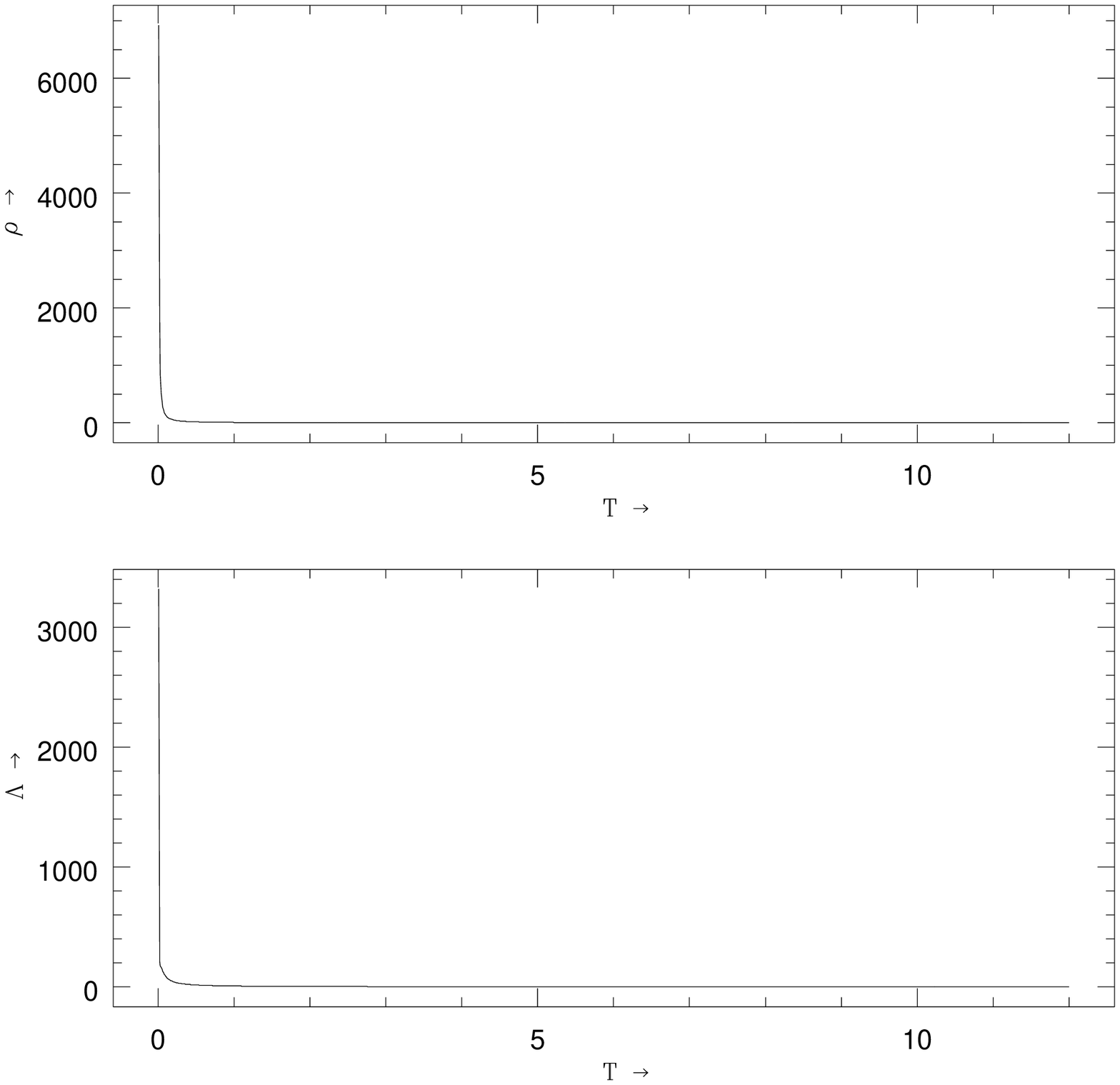}}
%%\centerline{\psfig{file=fig2.eps,width=8cm}} %100 percent
\vspace*{13pt}
\caption{(Top) Plot of $\rho \rightarrow T$ and (bottom) $\Lambda \rightarrow
T$ for parameters $n=0.45$, $K=2.0$,  $k_3 = 1$, $\gamma = 0.5$, $\alpha=1$, $\xi_{0} = 1$ and
rest of the constants are set to 1.}
\label{fig3}
\end{figure}
%%%%%%%%%%%%%%%%%%%%%%%%%%%%%%%%%%%%%%%%%%%%%%%%%%%%%%%%%%%%%%%%%%%%%%%%%%%%%%%%%%%%%%
%%%%%%%%%%%%%%%%%%%%%%%%%%%%%%%%% SUBSUBSECTION 3.3.2  %%%%%%%%%%%%%%%%%%%%%%%%%%%%%%%%%
\subsubsection{Model II: ~ ~ Solution for $w = 1$}
When $w = 1$, Eq. (\ref{eq28}) reduces to $\xi = \xi_{0} \rho$. Hence in this case 
Eq. (\ref{eq58}), with the use of (\ref{eq59}), (\ref{eq27}) and (\ref{eq28}), leads to
\[
8\pi \rho = \frac{1}{[(1 + \gamma) - 3\xi_{0}D_{5}{D_{4}}^{-\frac{3}{2}}]} \Big[-\ell^{2}(11n^{2} - 4n - 9)
T^{-3n} + 16 n(n - 1)\ell k_{3}T^{\ell_{1} + n}
\]
\begin{equation}
\label{eq62}
+ n(n + 2){k_{3}}^{2}T^{2(n - 1)}\Big]{D_{4}}^{-3} - \alpha^{2}T^{-n}{D_{3}}^{-1}. 
\end{equation}
Eliminating $\rho(t)$ from Eqs. (\ref{eq59}) and (\ref{eq62}), we obtain 
\[
8\pi \Lambda = \frac{1}{[(1 + \gamma) - 3\xi_{0}D_{5}{D_{4}}^{-\frac{3}{2}}]} \Big[\{-\ell^{2}(11n^{2} 
- 4n - 9)T^{-3n} + 16 n(n - 1)\ell k_{3}T^{\ell_{1} + n}
\]
\[
+ n(n + 2){k_{3}}^{2}T^{2(n - 1)}\}{D_{4}}^{-3} - \alpha^{2}T^{-n}{D_{3}}^{-1}\Big] - 
\]
\begin{equation}
\label{eq63}
\left[\{-\frac{1}{2}(11n^{2} - 4n - 14)\ell^{2}T^{-3n} - 3n(2 - n)\ell k_{3}T^{\ell_{1} + n} + 
3n^{2}{k_{3}}^{2}T^{2(n - 1)}\}{D_{4}}^{-3} + 3\alpha^{2}T^{n}{D_{4}}^{-1}\right]
\end{equation}
In Mode I, from Eqs. (\ref{eq60}) and (\ref{eq61}), we observe that the energy density $\rho(t)$ 
and cosmological constant ($\Lambda(t)$) are decreasing functions of time (see Figure 2). The energy 
density is always positive. We also observe that the value of $\Lambda$ is small and positive at late 
times which is supported by recent type Ia supernovae observations \cite{ref9} $-$ \cite{ref12}. 
From Eqs. (\ref{eq62}) and (\ref{eq63}), we observe that Model II has the similar behaviour as 
Model I, so it is not reproduced here. \\

The scale of expansion and the shear are obtained as
\begin{equation}
\label{eq64}
\theta = 3\left[\frac{\ell(n - 2)}{2} T^{-3n/2} + k_{3} n T^{n - 1}\right]
\left(k_{3} T^{n} - 2\ell T^{\ell_{1}}\right)^{-3/2},
\end{equation}
\begin{equation}
\label{eq65}
\sigma = \frac{kT^{-3n/2}}{2}\left(k_{3} T^{n} - 2\ell T^{\ell_{1}}\right)^{-3/2}.
\end{equation}
From (\ref{eq64}) and (\ref{eq65}), we get
\begin{equation}
\label{eq66}
\frac{\sigma}{\theta} = \frac{k}{6}\left[k_{3} nT^{-\ell_{1} + n} + \frac{\ell (n - 2)}
{2}\right]^{-1}.
\end{equation}
In subcase $\Lambda = 0$ and $\xi = 0$, metric (\ref{eq57}) with expressions $p$, $\rho$, $\theta$ 
and $\sigma$ for this model are same as that of solution (\ref{eq34}) of 
Camci {\it et al.} \cite{ref40}. If we set $\xi_{0} = 0$, this metric (\ref{eq57}) represents the 
solution obtained by Pradhan {\it et al.} \cite{ref41}.

In subcase $\Lambda = 0$, $\xi = 0$, $n = 0$, after an inverse transformation, metric (\ref{eq57})
reduces to the form
\begin{equation}
\label{eq67}
ds^{2} = dt^{2} - K_{5}(t + t_{4})^{2/3}dx^{2} - e^{2\alpha x}\left[dy^{2} + K^{2}_{5}
(t + t_{4})^{4/3}dz^{2}\right],
\end{equation}
where $t_{4}$ is an integrating constant. This model is different from the model 
(\ref{eq36}) by a change of scale. 

In subcase $\Lambda = 0$, $k = 0$, same model as (\ref{eq37}) is obtained. \\

Further in subcase  $\Lambda = 0$, $\xi = 0$ $k_{3} = 0$, we see that the metric (\ref{eq57})
takes the form
\begin{equation}
\label{eq68}
ds^{2} = dt^{2} - 2\ell K_{6}(t + t_{5})^{2/3}\left[dx^{2} + e^{2\alpha x}
\left(b t^{m_{2}}dy^{2} + b^{-1}t^{-m_{2}} dz^{2}\right)\right],
\end{equation}
where $t_{5}$ is a constant, $bt^{m_{2}} = 2\ell K^{\frac{2 - 5n}{2 - n}}_{6}
(t + t_{5})^{\frac{2(2 - 5n)}{3(2 - n)}}$ and $K_{6} = \left[\frac{3(2 - n)}
{4\sqrt{2 \ell}}\right]^{2/3}$. This metric is only different from (\ref{eq37}) by
a change of sign. Also, in each of subcase the physical and kinematical properties
of obtained metric are same as that of Case(I). Therefore, we do not consider here them. \\

%%%%%%%%%%%%%%%%%%%%%%%%%%% SUBSECTION 3.4  %%%%%%%%%%%%%%%%%%%%%%%%%%%%%%%%%%
\subsection{Case (IV) : Let $B$ = $\tilde{\tau}^{n}$, where $n$ is any real number.} 
In this case Eq. (\ref{eq21}) gives
\begin{equation}
\label{eq69}
C = k_{4} \tilde{\tau}^{n}\exp{\left(\frac{k}{\ell_{1}} \tilde{\tau}^{\ell_{1}}
\right)}
\end{equation}
and then from (\ref{eq14}), we obtain
\begin{equation}
\label{eq70}
A^{2} = k_{4}\tilde{\tau}^{2n}\exp{\left(\frac{k}{\ell_{1}} \tilde{\tau}^{\ell_{1}}
\right)} 
\end{equation}
Hence the metric (\ref{eq4}) reduces to
\[
ds^{2} = \tilde{\tau}^{2n}\exp{\left(\frac{k}{\ell_{1}} \tilde{\tau}^{\ell_{1}}\right)}
\left[\tilde{\tau}^{n}\exp{\left(\frac{2k}{\ell_{1}} \tilde{\tau}^{\ell_{1}}\right)}
- dx^{2}\right]
\]
\begin{equation}
\label{eq71}
- e^{2 \alpha x}\left[dy^{2} + \exp{\left(\frac{2k}{\ell_{1}} \tilde{\tau}^{\ell_{1}}
\right)} - dz^{2}\right],
\end{equation}
where, without any loss of generality, the constant $k_{4}$ is taken equal to $1$. 
Expressions for physical and kinematical parameters for the model (\ref{eq71}) are 
not given here, but it is observed that the properties of the metric (\ref{eq71}) 
are same as that of the solution (\ref{eq42}), i.e. the Case (II). \\

%%%%%%%%%%%%%%%%%%%%%  SECTION 4  %%%%%%%%%%%%%%%%%%%%%%%%%%%%%%%%%%%%%%%%%%
\section{Concluding remarks} 

In this paper we have described a new exact solutions of EFES for Bianchi type V 
spacetime with a bulk viscous fluid as the source of matter and cosmological term $\Lambda$ 
varying with time. Using a generation technique followed by Camci {\it et al.} \cite{ref40},
it is shown that the Einstein's field equations are solvable for any arbitrary cosmic 
scale function. Starting from particular cosmic functions, new classes of spatially
homogeneous and anisotropic cosmological models have been investigated for which the
fluids are acceleration and rotation free but they do have expansion and shear. For
$\alpha = 0$ in the metric (\ref{eq4}), we obtained metrics as LRS Bianchi type I model
(Hajj-Boutros \cite{ref33}, Hajj-Boutros and Sfeila \cite{ref34}, Ram \cite{ref35}, 
Mazumder \cite{ref36}, Pradhan and Kumar \cite{ref37}) and Pradhan {\it et al.} \cite{ref41}. 
It is also seen that the solutions obtained by Camci {\it et al.} \cite{ref40}, Ram \cite{ref35}, 
Pradhan and Kumar \cite{ref37} and Bianchi type V models studied by Pradhan {\it et al.} \cite{ref41} 
are particular cases (except one) of our solutions. \\

The cosmological constants in all models given in Section $3$ are decreasing functions
of time (except one, Model II of Case (I)) and they all approach a small positive value at 
late times which are supported by the results from recent supernova observations recently 
obtained by the High-z Supernova Team and Supernova Cosmological Project 
(Garnavich {\it et al.} \cite{ref9}, Perlmutter {\it et al.} \cite{ref10}, 
Riess {\it et al.} \cite{ref11}, Schmidt {\it et al.}\cite{ref12}). \\

The features of these new solutions are that from our wide range of choice of parameters for Model I 
of Cases (I), (II), and (III), it is apparent from figures that energy density $\rho(t)$ and dynamic 
cosmological term $\Lambda(t)$ are decreasing function of time, remain positive and small at later 
stage during evolution. These two quantities remain finite and do not become zero at later stage. 
Hence this seem to be physically viable to explore physical mechanism for further detail study depending 
on relevance of the physical problem. These studies show that the bulk viscous effect is apparent 
on $\Lambda(t)$. In many astrophysical situations the viscosity calculated using statistical 
consideration do not give satisfactory viscous number. Hence the specific thing about bulk viscosity 
will depend on the details of the viscous nature of the matter. We have explored Model II of Cases (I), 
(II) and (III). So we feel that it is not necessary to display behaviour of $\rho(t)$ and $\Lambda(t)$. 
We have studied these models and no anomaly has been observed. \\

Model II of Case (I) is explored for variety of parameters. We find that many cases energy density 
decreases sharply and becomes negative very fast and later stage also remains negative but there is 
a slight increase at later stage and remains constant (negative constant), which we feel undesirable 
feature. So it is not discussed in the paper. Also in many cases it initially oscillatory. In this 
case $\Lambda(t)$ shows very peculiar behaviour. Only in few cases $\Lambda$ is initially negative then 
becomes positive but in all cases during initial evolution $\Lambda$ has singularities near $t$ close 
to zero and a few at later finite time. The cause of this behaviour is not very apparent from the model.
Otherwise in wide range of parameters $\Lambda$ shows eratic behaviour. Most of the cases singular and 
eratic behaviour of $\Lambda(t)$ is persistent and cause is not known. So we do not discuss further. 
This may require detailed study separately and this paper is not appropriate place.  

\section*{Acknowledgements}
The authors would like to thank the Inter-University Centre for Astronomy and Astrophysics, Pune, 
India for providing facility where this work was carried out. 
\newline
\newline
%% \newpage
%\nonumsection{References}

\end{document}